\documentclass[aps,superscriptaddress,twocolumn]{revtex4-1}

\usepackage{amsbsy}
\usepackage{amsfonts}
\usepackage{amssymb}
\usepackage{amsmath}
\usepackage{color}
\usepackage{graphicx}
\usepackage{xspace}
\usepackage[normalem]{ulem}
\usepackage{stmaryrd}
\usepackage{natbib}

\usepackage{dsfont}
\newcommand{\idop}{\mathds{1}}

\newcommand{\bb}[0]{\begin{eqnarray}}
\newcommand{\ee}[0]{\end{eqnarray}}
\newcommand{\ket}[1]{| #1 \rangle}
\newcommand{\bra}[1]{\langle #1 |}
\newcommand{\moy}[1]{\ensuremath{\langle #1\rangle}\xspace}

\begin{document}

\title{Probing quantum fluctuation theorems in engineered reservoirs}

\author{C. Elouard}
\affiliation{CNRS and Universit\'e Grenoble Alpes, Institut N\'eel, F-38042 Grenoble, France}

\author{N. K. Bernardes}
\affiliation{Departamento de F\'isica, Universidade Federal de Minas Gerais, Belo Horizonte, Caixa Postal 702, 30161-970, Brazil}

\author{A. R. R. Carvalho}
\affiliation{Centre for Quantum Dynamics, Griffith University, Nathan, Queensland 4111, Australia}

\author{M. F. Santos}
\affiliation{Instituto de F\'isica, Universidade Federal do Rio de Janeiro, Caixa Postal 68528, Rio de Janeiro, RJ 21941-972, Brazil}

\author{A. Auff\`eves}
\email{alexia.auffeves@neel.cnrs.fr}
\affiliation{CNRS and Universit\'e Grenoble Alpes, Institut N\'eel, F-38042 Grenoble, France}

\begin{abstract}
Fluctuation Theorems are central in stochastic thermodynamics, as they allow for quantifying the irreversibility
of single trajectories. Although they have been experimentally checked in the classical regime, a practical demonstration in the framework of quantum open systems is still to come. Here we propose a realistic platform to probe fluctuation theorems in the quantum regime. It is based on an effective two-level system coupled to an engineered reservoir, that enables the detection of the photons emitted and absorbed by the system. When the system is coherently driven, a measurable quantum component in the entropy production is evidenced. We quantify the error due to photon detection inefficiency, and show that the missing information can be efficiently corrected, based solely on the detected events. Our findings provide new insights into how the quantum character of a physical system impacts its thermodynamic evolution.
\end{abstract}

\pacs{Valid PACS appear here}

\maketitle

\section{Introduction}
The existence of some preferred direction of time is a fundamental concept of physics, captured by the second Law of thermodynamics \cite{Eddington,Lebowitz93}. The irreversibility of physical phenomena is manifested by a strictly positive entropy production, classically identified with the change of entropy of the closed system considered. In the textbook situation of a system ${\cal S}$ exchanging work $W$ with some external operator and heat $Q_\text{cl}$ with a thermal reservoir ${\cal R}$ of temperature $T$, entropy production equals the entropy change of the system $\Delta S_\text{sys}$ and the reservoir $\Delta S_\text{res} = -Q_\text{cl}/T$, and reads

\begin{equation}\label{DiS}
\Delta_\text{i}S =(W - \Delta F)/T.
\end{equation}

\noindent $\Delta F$ is the change of the system's free energy, satisfying $\Delta S_\text{sys} = (W+Q_\text{cl}-\Delta F)/T$.

\begin{figure}
\begin{center}
\includegraphics[width=0.45\textwidth]{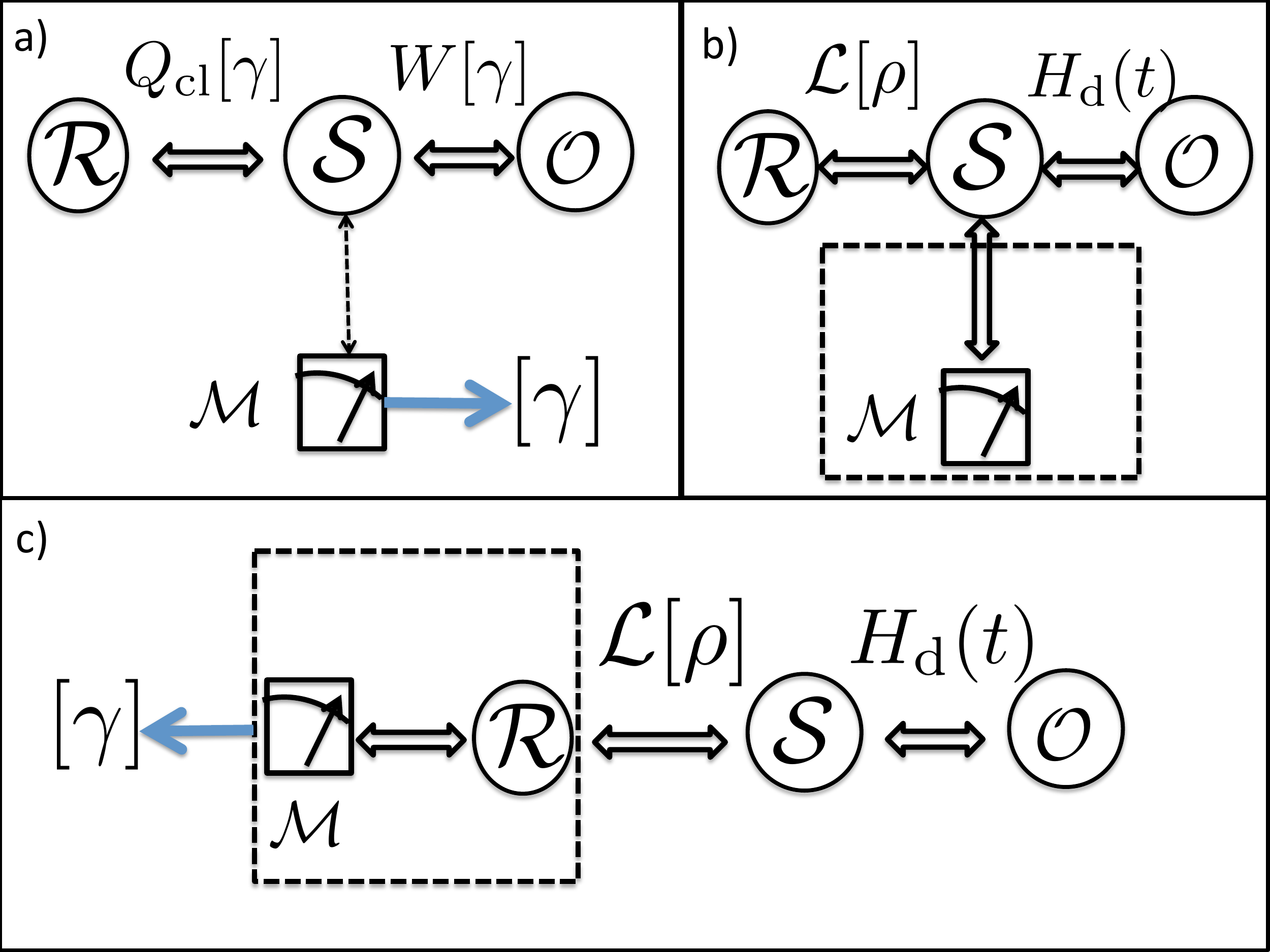}
\end{center}
\caption{a) Classical setup to test FT on a system $\cal S$ exchanging heat with a thermal bath $\cal R$ and work with an external operator ${\cal O}$ . The stochastic amounts of heat $Q_\text{cl}[\gamma]$ and work $W[\gamma]$ are reconstructed by recording the classical trajectory $\gamma$ of the system. b) ${\cal S}$ is a quantum system driven through a Hamiltonian $H_\text{d}(t)$ and interacting with a heat bath through the Lindbladian ${\cal L}[\rho]$. Monitoring ${\cal S}$ can randomly perturb its evolution and corresponds to an additional dissipative channel. c) Proposed strategy: Monitoring a reservoir engineered to simulate a thermal bath. \label{fig1} }
\end{figure}

Focusing on the transformations of microscopic systems, stochastic thermodynamics has extended the concepts of thermodynamics to single stochastic realizations or  ``trajectories" of the system in its phase space \cite{Seifert05,Sekimoto10,Jarzynski11}. In this framework, heat and work exchanges become stochastic quantities defined for single trajectories $\gamma$, which can give rise to a negative entropy production $\Delta_\text{i}s[\gamma] <0$. The fluctuations of $\Delta_\text{i}s[\gamma] $ verify the central Fluctuation Theorem (FT) \cite{Seifert08} $\langle e^{-\Delta_\text{i}s[\gamma]/k_\text{B}} \rangle_\gamma=1$, while the Second Law of thermodynamics remains valid on average. A famous example of such central FT is provided by Jarzynski's equality (JE) \cite{Jarzynski97,Jarzynski11} characterizing isothermal transformations of systems driven out of equilibrium. Eq.\eqref{DiS} remains valid at the trajectory level, such that entropy produced along a single trajectory $\gamma$ verifies $\Delta_\text{i}s[\gamma] =(W[\gamma]- \Delta F)/T$.  $W[\gamma]$ is the stochastic work exchange for the trajectory $\gamma$, yielding JE:
\begin{equation}
\langle e^{-W[\gamma] / k_\text{B}T} \rangle_\gamma = e^{-\Delta F/ k_\text{B}T} \label{JE}
\end{equation} 

\noindent In the classical regime, the stochastic work exchanges $W[\gamma]$ are inferred from the knowledge of the trajectory, which can be recorded without perturbing the system in principle with an arbitrary precision (Fig.\ref{fig1}a). Using this protocol, JE has been experimentally probed in various setups, e.g. with Brownian particles \cite{Toyabe10,Berut12} or electronic systems \cite{Saira12}. \\
 
The recent developments of quantum and nano-technologies have urged to explore the validity of FTs for out-of-equilibrium quantum systems. In the particular case of Jarzynski's protocol, the system coupled to a thermal reservoir can now be coherently driven. However in the quantum world, monitoring a system perturbs its evolution (Fig.\ref{fig1}b), posing severe difficulties to measure and even to define work exchanges. In a series of pioneering papers \cite{Talkner07,Talkner09,Esposito09,Campisi11} it was shown that JE is still valid and can be probed, provided that the system's (resp. reservoir's) internal energy change $ \Delta U[\gamma]$  (resp. $- Q_\text{cl}[\gamma]$) can be known from projective energy measurements performed in the beginning and at the end of the transformation. The work is then inferred from the First Law: $\Delta U[\gamma] = W[\gamma] + Q_\text{cl}[\gamma]$. Known as the two-points measurement scheme, this seminal extension of JE in the quantum world has triggered numerous investigations, motivated by two major challenges. The first challenge is practical: How to measure energy changes induced by a microscopic system on a large reservoir? One possible strategy relies on finite size reservoirs whose temperature is sensitive to heat exchanges with the system \cite{Pekola13, Hekking13,Campisi14}, such that both emission and absorption events can be detected using fine calorimetry. But such reservoirs are by essence non Markovian, therefore affecting the dynamics of the system and the resulting work distribution \cite{Suomela16}. 

Another strategy involves the convenient platforms provided by superconducting circuits and semi-conducting quantum photonics. In both cases, the radiation produced by quantum emitters (superconducting qubits or quantum dots) is efficiently funneled into well-designed waveguides, and is thus recorded with high efficiency. This recent experimental ability has lead to the development of bright single photon sources \cite{Claudon10, Somaschi16} and to the monitoring of quantum trajectories of superconducting quantum bits \cite{Riste13,Murch13,Weber14}. However, standard schemes based on photo-counters do not allow for the recording of photons absorbed by the system, a major drawback for quantum thermodynamics purposes. 

The second challenge is of fundamental nature, and consists in identifying quantum signatures in quantum FTs. For instance, the coupling to a thermal bath erases quantum coherences during the system's evolution. This is an irreversible process, expected to induce some genuinely quantum entropy production \cite{Auffeves15,Elouard16}. Furthermore, the incoherent energy exchange between the system and the bath (heat) is carried by quantized excitations. Thus any information depending on the recording of these excitations may be significantly hindered by the eventual inefficiency of the detection scheme. So far, an experimental observation of such genuinely quantum effects has remained elusive.
 
\begin{figure}
\begin{center}
\includegraphics[width=0.45\textwidth]{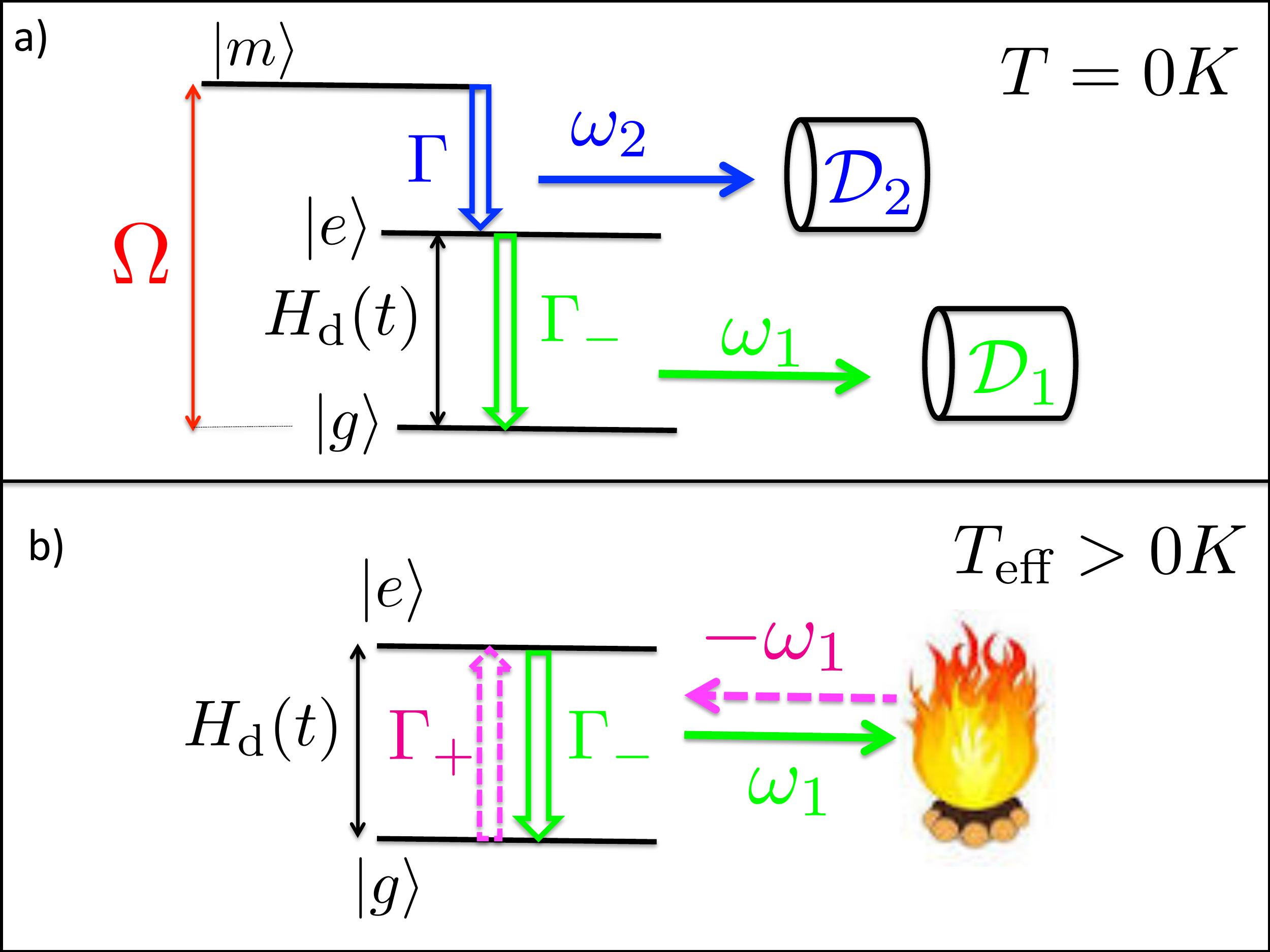}
\end{center}
\caption{a) Engineered environment based on a three-level atom. The state $\ket{m}$ has a very short lifetime due to spontaneous emission in vacuum characterized by the rate $\Gamma$. The transition $\ket{e}$ to $\ket{g}$ (resp. $\ket{m}$ to $\ket{e}$) is associated with emission of a photon of energy $\omega_1$ (resp. $\omega_2$) detected by ${\cal D}_1$ (resp. ${\cal D}_2$). The transition $\ket{e} - \ket{g}$ is driven by the Hamiltonian $H_\text{d}(t)$.
b) Equivalent description in terms of an effective thermal bath. In the presence of a weak classical drive of Rabi frequency $\Omega$ resonant with the transition $\ket{m} \rightarrow \ket{e}$, level $\ket{m}$ can be adiabatically eliminated. This result in an effective incoherent rate $\Gamma_+ = 4\Omega^2/\Gamma$ associated with transition $\ket{g} \rightarrow \ket{e}$ (dotted arrow), simulating a qubit coupled to an effective thermal bath of temperature $T_\text{eff} = \hbar\omega_1/k_\text{B}\log(\Gamma_-/\Gamma_+)$. \label{fig2} }
\end{figure}

Here we propose a strategy that addresses both the practical and fundamental issues raised above (Fig.\ref{fig1}c and Fig.\ref{fig2}). Our proposal is based on an effective two-level system (TLS) coupled to an engineered thermal bath already introduced by some of us \cite{Carvalho11,Santos11} to perform incoherent quantum computing \cite{Santos12}. A great interest of this scheme is that it allows for the detection of the emission and the absorption of photons in the reservoir, while still keeping its Markovian character.  Eventually, both the heat exchanges and the quantum trajectory followed by the system can be fully reconstructed. Note that a quite similar strategy is proposed in \cite{Horowitz12} to access quantum trajectories of a forced Harmonic oscillator.

We first introduce and validate the platform by checking JE in the simple case where the external drive does not create any coherences in the system, and heat exchanges are perfectly detected. We then turn to the case of a coherence-inducing drive for which we provide a full thermodynamic analysis. Genuinely quantum signatures in thermodynamic quantities are evidenced, as well as how to measure them. Finally, we investigate the influence of the detectors finite efficiency on the expression of the measured FTs. The non-detected heat results in a modified expression of JE that includes an extra term in the entropy production. We show that this term is related to the measurement record and can be computed from the available data.\\

\section{Engineered thermal bath}

We consider a system whose three levels are denoted by $\ket{m}$, $\ket{e}$ and $\ket{g}$ of respective energy $E_m>E_e>E_g$ (see Fig.\ref{fig2}a). This three-level system is coupled to an electro-magnetic reservoir at zero temperature, such that $\ket{m}$ decays to state $\ket{e}$ (resp. $\ket{e}$ decays to $\ket{g}$) with a spontaneous emission rate $\Gamma$ (resp. $\Gamma_-$). We assume that $\ket{m}$ is a metastable level, such that $\Gamma \gg \Gamma_-$. We denote $\omega_1 = (E_e-E_g)/\hbar$ and $\omega_2 = (E_m-E_e)/\hbar$. In addition, the atom is weakly driven by a laser resonant with transition $\omega_1+\omega_2$ and of Rabi frequency $\Omega$, satisfying $\Omega \ll \Gamma$. Due to its short lifetime, level $\ket{m}$ can be adiabatically eliminated, resulting in an effective incoherent transition rate $\Gamma_+ = 4\Omega^2/\Gamma$ from state $\ket{g}$ to state $\ket{e}$ \cite{Carvalho11}. The states $\ket{e}$ and $\ket{g}$ define our effective qubit of interest (Fig.\ref{fig2}b). Without external drive, this qubit relaxes towards some effective thermal equilibrium characterized by the temperature $T_\text{eff}$,  satisfying $e^{-\hbar \omega_1/k_\text{B}T_\text{eff}} = \frac{\Gamma_+}{\Gamma_-}$. 

In what follows we drive the qubit transition with a Hamiltonian $H_\text{d}(t)$, such that $\omega_1$ may depend on time: the rates $\Gamma_{\pm}$ can be adjusted accordingly to keep the effective temperature constant. These adjustments are discussed in Ref.\cite{Carvalho11} and the Methods Section at the end of this manuscript. The dynamics of the effective qubit density matrix $\rho$ is ruled by the master equation:
\begin{equation}
\dot \rho = -\dfrac{i}{\hbar}\left[H_0+H_\text{d}(t),\rho\right] + \Gamma_+ \mathcal{L}[\sigma^\dagger]\rho + \Gamma_- \mathcal{L}[\sigma]\rho,\label{Meq}
\end{equation}

\noindent where $H_0 = (\hbar \omega_1/2) \sigma_z$ with $\sigma_z = \ket{e}\bra{e}-\ket{g}\bra{g}$. We have introduced the dissipation super-operator $\mathcal{L}[X]\rho = X\rho X^\dagger - \tfrac{1}{2}\{X^\dagger X,\rho\}$, with $\sigma = \ket{g}\bra{e}$. Assuming that a photo-counter ${\cal D}_1$ (resp. ${\cal D}_2$) detects the photons emitted at frequency $\omega_1$ (resp. $\omega_2$), one can formulate the evolution of the qubit state conditioned to the measurement records of the detectors in terms of quantum jumps \cite{WisemanBook}. Assuming an initial known pure state $\ket{\Psi_0}$ of the qubit, and discretizing the time between $t_\text{i}$ and $t_\text{f}$ such as $t_n = t_\text{i} + ndt$ (with $n\in \llbracket 0,N \rrbracket$ and $t_N = t_\text{f}$), the evolution of the system features a stochastic trajectory $\gamma$ of pure states $\ket{\psi_\gamma(t_n)}$. The trajectory is generated by applying a sequence of operators $M_{{\cal K}(t_n)}$, where ${\cal K}(t_n)$ stands for the stochastic measurement outcome at time $t_n$. 

Namely, detecting a photon on detector ${\cal D}_1$ (resp. ${\cal D}_2$) at time $t_n$ corresponds to applying the operator $M_1 = \sqrt{\Gamma_-dt}\sigma$ (resp. $M_2 = \sqrt{\Gamma_+dt}\sigma^\dagger$) on the qubit state $\ket{\psi_\gamma(t_n)}$: remarkably in this scheme, the absorption of a photon from the effective heat bath (Fig.\ref{fig2}b) is detectable and actually corresponds to the emission of a photon of frequency $\omega_2$ (Fig.\ref{fig2}a). If no photon is detected, the no-jump 
operator $M_0 = \idop - idt H(t) - \tfrac{1}{2} M_1^\dagger M_1^{}- \tfrac{1}{2} M_2^\dagger M_2^{}$ is applied. The qubit state is then renormalized. Each of these possible evolutions occurs at time $t_n$ with probabilities
\bb
p_{{\cal K}(t_n)} = \bra{\psi_\gamma(t_n)}M_{{\cal K}(t_n)}^\dagger M_{{\cal K}(t_n)} \ket{\psi_\gamma(t_n)},\quad {\cal K} \in \{0,1,2\}.\nonumber\\
\ee

\section{Stochastic thermodynamic quantities}
\label{Thermo}
We now define the thermodynamic quantities associated with the system's quantum trajectory $\ket{\psi_\gamma(t_n)}$. Following \cite{Elouard16}, the internal energy of the qubit at time $t_n$ is defined as $U_\gamma(t_n) =  \bra{\psi_\gamma(t_n)}H(t_n)\ket{\psi_\gamma(t_n)}$, with $H(t) = H_0+H_\text{d}(t)$ the total Hamiltonian of the system. The variation of this quantity along trajectory $\gamma$ splits into three terms: there is a work increment $\delta W_\gamma(t_n) $ due to the Hamiltonian time-dependence and given by:
\bb
\delta W_\gamma(t_n) = \bra{\psi_\gamma(t_n)}dH_\text{d}(t_n)\ket{\psi_\gamma(t_n)}, \quad {\cal K}(t_n)=0.\label{dW}
\ee 

\noindent There is also a classical heat $Q_\text{cl}[\gamma]$ corresponding to the energy exchanged with the effective thermal bath under the form of emitted or absorbed excitations of energy $\hbar \omega_1(t_n)$, where $\omega_1(t_n)$ is the qubit effective transition frequency defined as  $\omega_1(t) = (\bra{e}H(t)\ket{e}-\bra{g}H(t)\ket{g})/\hbar$. This increment reads
\bb
\delta Q_{\text{cl},\gamma}(t_n) = \left\{\begin{array}{c}
-\hbar\omega_1(t_n),\quad {\cal K}(t_n)=1 \\ 
\hbar\omega_1(t_n),\quad{\cal K}(t_n)=2.\label{dQcl}
\end{array} \right.
\ee

\noindent Finally, a third contribution must be introduced whenever the system's dynamics creates coherences in the $\{ \ket{e}; \ket{g} \}$ basis. Defined as quantum heat in \cite{Elouard16}, its increment when a jump takes place reads:
\bb
\delta Q_{\text{q},\gamma}(t_n) = \left\{\begin{array}{c}
\hbar\omega_1(t_n)  \vert\langle g\ket{\psi_\gamma(t_n)}\vert^2,\quad {\cal K}_\gamma(t_n)=1 \\ 
-\hbar\omega_1(t_n)  \vert\langle e\ket{\psi_\gamma(t_n)}\vert^2,\quad{\cal K}_\gamma(t_n)=2,
\end{array} \right.\label{dQq}
\ee

\noindent while if no jump takes place, it reads

\begin{widetext}
\bb
\delta Q_{\text{q},\gamma}(t_n) &=& -\dfrac{1}2  \bra{\psi_\gamma(t_n)}\{M_1^\dagger M_1-p_1(t_n),H(t_n)\}\ket{\psi_\gamma(t_n)}\nonumber \\
 &&-\dfrac{1}2  \bra{\psi_\gamma(t_n)}\{ M_2^\dagger M_2 -p_2(t_n),H(t_n)\}\ket{\psi_\gamma(t_n)}
,\quad{\cal K}_\gamma(t_n)=0.
\ee
\end{widetext}

\noindent The first Law expressed for the trajectory $\gamma$ is then given by
\bb \label{1law}
\Delta U[\gamma] = U_\gamma(t_f) - U_\gamma(t_i)= W[\gamma] +  Q_\text{cl}[\gamma]+Q_\text{q}[\gamma], 
\ee 
\\
\noindent where $O[\gamma]$ denotes the integrated quantity over the whole trajectory. \\

\begin{figure*}[t]
\begin{center}

\includegraphics[width=0.95\textwidth]{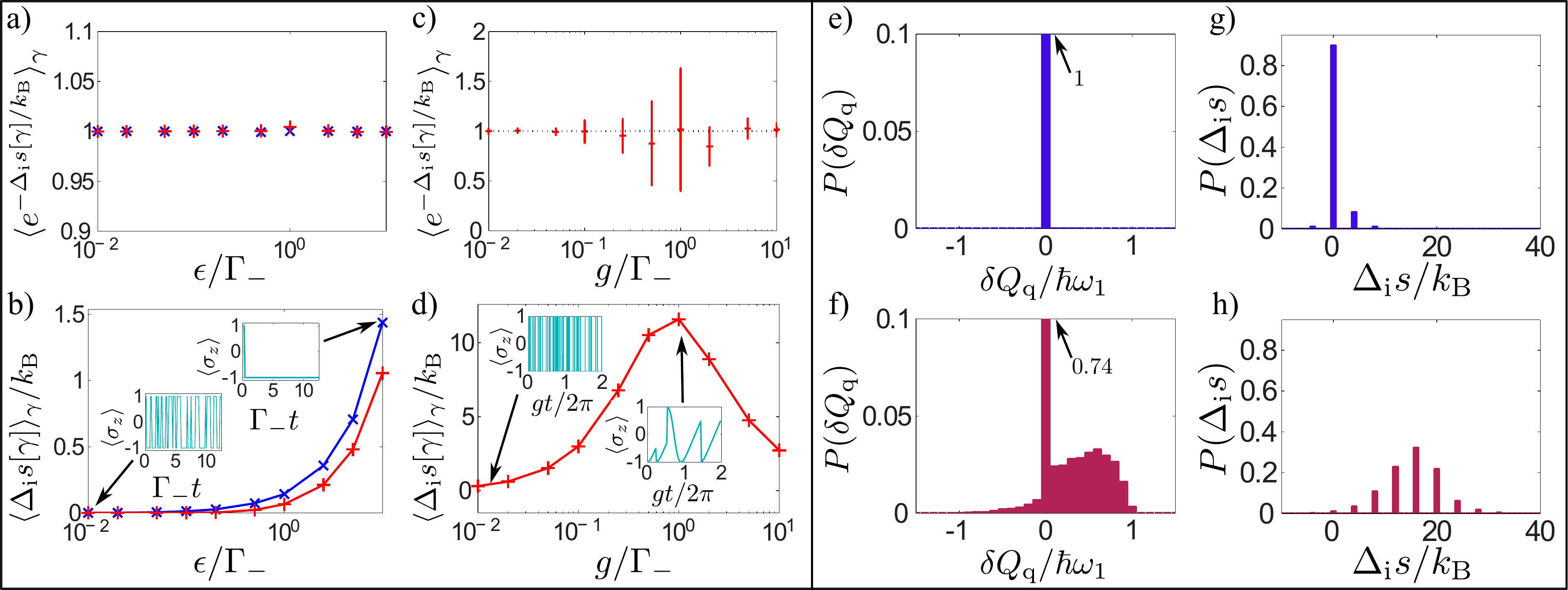}\\

\end{center}

\caption{a),c) Evolution of the parameter $\langle e^{-\Delta_\text{i}s[\gamma]/k_\text{B}}\rangle_\gamma$ and b),d) of the mean entropy production $\moy{\Delta_i s[\gamma]}_\gamma$ . We have considered the case of Landauer's drive $H_\text{d}(t) = (\hbar \omega_1/2) \epsilon (t-t_i) \sigma_z$ for a),b) and of a coherent drive $H_\text{d}(t) = (\hbar g/2)(\sigma e^{i\omega_1 t} + \sigma^\dagger e^{-i\omega_1 t})$ for c),d). Both quantities are studied as a function of the respective driving strengths $\epsilon/\Gamma_-$ and $g/\Gamma_-$. Insets: evolution of $\moy{\sigma_z} = \bra{\psi_\gamma(t)}\sigma_z\ket{\psi_\gamma(t)}$ along a single trajectory $\gamma$, in two different regimes of the studied transformation. In b) two different temperatures are used: $\hbar\omega_1/k_\text{B}T_\text{eff} =3$ (red `+') and $\hbar\omega_1/k_\text{B}T_\text{eff} =0.3$ (blue `x'). In d), $\hbar\omega_1/k_\text{B}T_\text{eff} =3$.  Distributions of the quantum heat increment e),f) and of the entropy produced along a single trajectory $\gamma$ g),h) in the case of a coherent drive for $g/\Gamma_- = 0.01$ (e,g) and $g/\Gamma_- = 1$ (f,h). {\it Parameters}: Number of trajectories $N_\text{traj} = \:10^{5}$ for a),b), $N_\text{traj} = 2\times 10^{6}$ for c),d), $N_\text{traj} = \:10^{3}$ for e) to h).} \label{fig3}
\end{figure*}

\section{Jarzynski equality in the quantum regime}
Jarzynski's protocol generally consists in preparing the qubit in the thermal equilibrium state $\rho_\text{i}=Z_\text{i}^{-1}e^{H(t_\text{i})/k_\text{B}T_\text{eff}}$, then performing a strong measurement of the qubit energy state to project it on an initial pure state $\ket{i}$ of internal energy $U_\text{i}$. The qubit is driven out of equilibrium until time $t_\text{f}$ when a final strong energy measurement is performed, projecting the qubit onto the final state $\ket{f}$ of internal energy $U_\text{f}$. Note that the final density matrix is in general different from the final thermal equilibrium state $Z_\text{f}^{-1}e^{H(t_\text{f})/k_\text{B}T_\text{eff}}$. We have introduced the initial and final partition functions $Z_\text{i}$ and $Z_\text{f}$. The two strong measurements allow inferring the qubit internal energy change $\Delta U[\gamma] = U_\text{f} - U_\text{i}$, while the bath monitoring allow recording the classical heat exchange $Q_\text{cl}$. As stated in the introduction (the demonstration is provided in Section \ref{FiniteDetEff}), the entropy produced in a single trajectory $\gamma$ equals $\Delta_\text{i} s[\gamma] = (\Delta U[\gamma] - Q_\text{cl}[\gamma]- \Delta F)/T_\text{eff}$ \cite{Horowitz13,Elouard16}. Inserting Eq.\eqref{1law}, it becomes 

\bb
\Delta_\text{i}s[\gamma] = \frac{1}{T_\text{eff}}(W[\gamma]-\Delta F)+ \frac{Q_\text{q}[\gamma]}{T_\text{eff}}. \label{eq:Si}
\ee 

\noindent While fully compatible with former expressions for entropy production, Eq.(\ref{eq:Si}) reveals two components respectively involving the amounts of work and quantum heat exchanged. The second component represents a genuinely quantum contribution due to the presence of quantum coherences in the qubit bare energy basis, that quantitatively relates entropy production and energetic fluctuations. In the following, we show that this quantum contribution can be measured in our setup.\\

We first validate our platform by considering as driving Hamiltonian $H_\text{d}(t) = (\hbar \omega_1/2) \epsilon \sigma_z (t-t_\text{i})$ for $t_\text{i}\leq t \leq t_\text{f}$. The work and heat exchanged in this case take particularly simple forms, the work being due to time-varying energy levels, and the heat to stochastic population changes induced by the bath. More precisely, we have $\delta W_\gamma(t_n) = \langle \sigma_z(t_n) \rangle \hbar \epsilon dt$ and $\delta Q_{\text{cl},\gamma}(t_n) = (\langle \sigma_z(t_{n+1}) \rangle-\langle \sigma_z(t_n)\rangle)  \hbar \omega_1(t_n)$, while the quantum heat increment and the quantum component of entropy production are zero. Note that this transformation boils down to the famous Landauer's protocol \cite{Landauer61,Berut12}. 
In Fig.\ref{fig3}a and b, we have plotted the quantity $\langle e^{-\Delta_\text{i}s[\gamma]/k_\text{B} } \rangle_\gamma$ and mean entropy production $\langle \Delta_\text{i}s[\gamma] \rangle_\gamma$ for different values of the coupling constant $\epsilon$ and temperatures. First note that, as expected, JE is verified for the thermodynamic quantities previously defined. Also note that the entropy production vanishes for $\epsilon \ll \Gamma_-$. This is also expected since, in this limit, the drive is quasi-static and the qubit is always in equilibrium with the heat bath. This situation corresponds to a reversible transformation and, in our example, amounts to a large exchange of photons between the system and the bath before the energy of the system significantly changes. Finally, entropy production drastically increases when $\epsilon \gtrsim \gamma$. In this case, very few photons are exchanged and the system is driven far from equilibrium. \\

We now turn to a case with no classical counterpart by considering a drive $H_\text{d}(t) = (\hbar g/2)(\sigma e^{i\omega_1 t} + \sigma^\dagger e^{-i\omega_1 t})$ for $t_\text{i} < t < t_\text{f}$ and $H_\text{d}(t_\text{i}) = H_\text{d}(t_\text{f}) = 0$. This can be implemented by coupling the atom with a classical light field resonant with the transition $\omega_1$. It gives rise to Rabi oscillations, i.e. to the reversible exchange of energy (work) between the qubit and the field under the form of the coherent and periodic emission and absorption of photons at the Rabi frequency $g$. Note that this scenario deviates from the previous one as, here, work exchanges induce coherences and population changes in the qubit bare energy basis. According to our definitions, the coupling of this coherently driven qubit to the bath will now give rise to both classical and quantum heat exchanges, as well as a quantum component in the entropy production.

We first check that JE is recovered from the detection events in this quantum regime, as it appears in Fig\ref{fig3}c. This shows that our generalized thermodynamic quantities are properly defined. A more thorough analysis can be drawn from
Fig.\ref{fig3}d, where we plot the average entropy production as a function of the ratio $g/\Gamma_-$. There, one can identify three different situations, two of which corresponding to reversible transformations ($\langle \Delta_\text{i} s \rangle_\gamma \rightarrow 0$). For a very weak drive $g\ll \Gamma_-$, the qubit is always in a thermal equilibrium state. This regime corresponds to reversible quasi-static isothermal transformations. In the other extreme, i.e. when $g\gg \Gamma_-$, the drive is strong enough to induce almost unperturbed Rabi oscillations. In this case, the transformation is too fast to allow for a stochastic event to take place and heat to be exchanged, i.e. it is adiabatic in the thermodynamic sense and once again reversible.

The intermediate scenario where $g\sim \Gamma_-$ gives rise to a maximal average entropy production. Note that, in this situation, the thermodynamic time arrow has a completely different nature from the classical case studied before. Here, entropy is mostly produced by the frequent interruptions of the qubit Rabi oscillations caused by the stochastic exchanges of photons with the bath (a.k.a. heat) (Inset of Fig\ref{fig3}d). During such quantum jumps, the quantum coherences induced by the driving process are erased, giving rise to the exchange of quantum heat, and to entropy production of quantum nature. The deep connection between quantum heat and entropy production is captured in Eq.\eqref{eq:Si} where they are quantitatively related, and in Fig\ref{fig3}(e to h), where both histograms concentrate around zero in the reversible regime, and take finite values in the irreversible regime. Such a connection between quantum irreversibility and energy fluctuations has already been evidenced in \cite{Elouard16} in a different context, where stochasticity is caused by quantum measurement instead of a thermal bath. In both cases, quantum heat exchanges and entropy production appear as two thermodynamic signatures of the same phenomenon, i.e. coherence erasure. \\

\section{Finite detection efficiency.}
\label{FiniteDetEff}
Up to now, we have assumed that the detection of heat is $100\%$ efficient, i.e. that all the photons emitted by the three level atom can be detected. In practice, state-of-the art devices only allow for a fraction $\eta$ of these photons to be indeed detected \cite{Natarajan12,Goltsman01}. A more realistic scenario is now investigated by considering that the photo-detectors have a finite efficiency $\eta$. As a consequence, between two detected photons, there exist several different possible (or fictitious) trajectories of pure states, which cannot be distinguished by the measurement record: Each of these trajectories corresponds to a particular sequence of undetected emissions and absorptions. This effect induces a decrease of the purity of the qubit state which has to be described by a stochastic density matrix $\rho_\gamma$, rather than a wavevector. In the limit where $\eta = 0$ (no detector), the evolution of this density matrix is captured by Eq.\eqref{Meq}, such that $\rho_\gamma(t) = \rho(t)$.

For $0<\eta<1$, the evolution of $\rho_\gamma(t)$ is still conditioned on the stochastic measurement outcome of the detector ${\cal K}_\gamma(t_n)$, that now corresponds to applying a set of super-operators $\{{\cal E}^{(\eta)}_{\cal K}[\rho_\gamma]\}$ (${\cal K}\in\{0,1,2\}$). Detecting one photon in detector ${\cal D}_1$ (resp. ${\cal D}_2$) between time $t_n$ and $t_n+dt$ corresponds to applying the super-operator ${\cal E}^{(\eta)}_{1}[\rho_\gamma] = \eta M_1 \rho_\gamma M_1^\dagger$ (resp. ${\cal E}^{(\eta)}_{2}[\rho_\gamma] = \eta M_2 \rho_\gamma M_2^\dagger$), which occurs with probability $p_1^{(\eta)}(t) = \eta \text{Tr}\{M_1^\dagger M_1\rho_\gamma(t)\}$ (resp. $p_2^{(\eta)}(t) = \eta \text{Tr}\{M_2^\dagger M_2\rho_\gamma(t)\}$). When no photon is detected, which occurs with probability $p_0^{(\eta)} = 1-p_1^{(\eta)}-p_2^{(\eta)}$, a super-operator decreasing the state purity is applied: 

\bb
&{\cal E}^{(\eta)}_{0}[\rho_\gamma] &= \idop -idt[H(t),\rho_\gamma(t)] \nonumber\\
&& - \dfrac{\eta dt}{2}(\Gamma_-\left\{\sigma^\dagger \sigma,\rho_\gamma(t)\right\} + \Gamma_+\left\{\sigma \sigma^\dagger,\rho_\gamma(t)\right\})\nonumber\\
&&+(1-\eta)\left(\Gamma_-dt \mathcal{L}[\sigma]+  \Gamma_+dt\mathcal{L}[\sigma^\dagger]\right)\rho_\gamma(t).
\label{E0eta}\ee

\noindent Note that ${\cal E}_0^{(\eta)}$ is a linear interpolation between ${\cal E}_0^{(1)}[\rho] = M_0 \rho_\gamma M_0^\dagger$ applied in the perfect efficiency quantum jump formalism and ${\cal E}_0^{(0)}[\rho]$ which is the Lindbladian (right-hand terms in Eq.\eqref{Meq}). After applying the super-operator ${\cal E}^{(\eta)}_{\cal K},$ the density matrix has to be divided by $p_{\cal K}^{(\eta)}$ in order to be renormalized. \\

\subsection*{Generalized Jarzynski equality.}
\begin{figure*}[t]
\begin{center}
\includegraphics[width=0.9\textwidth]{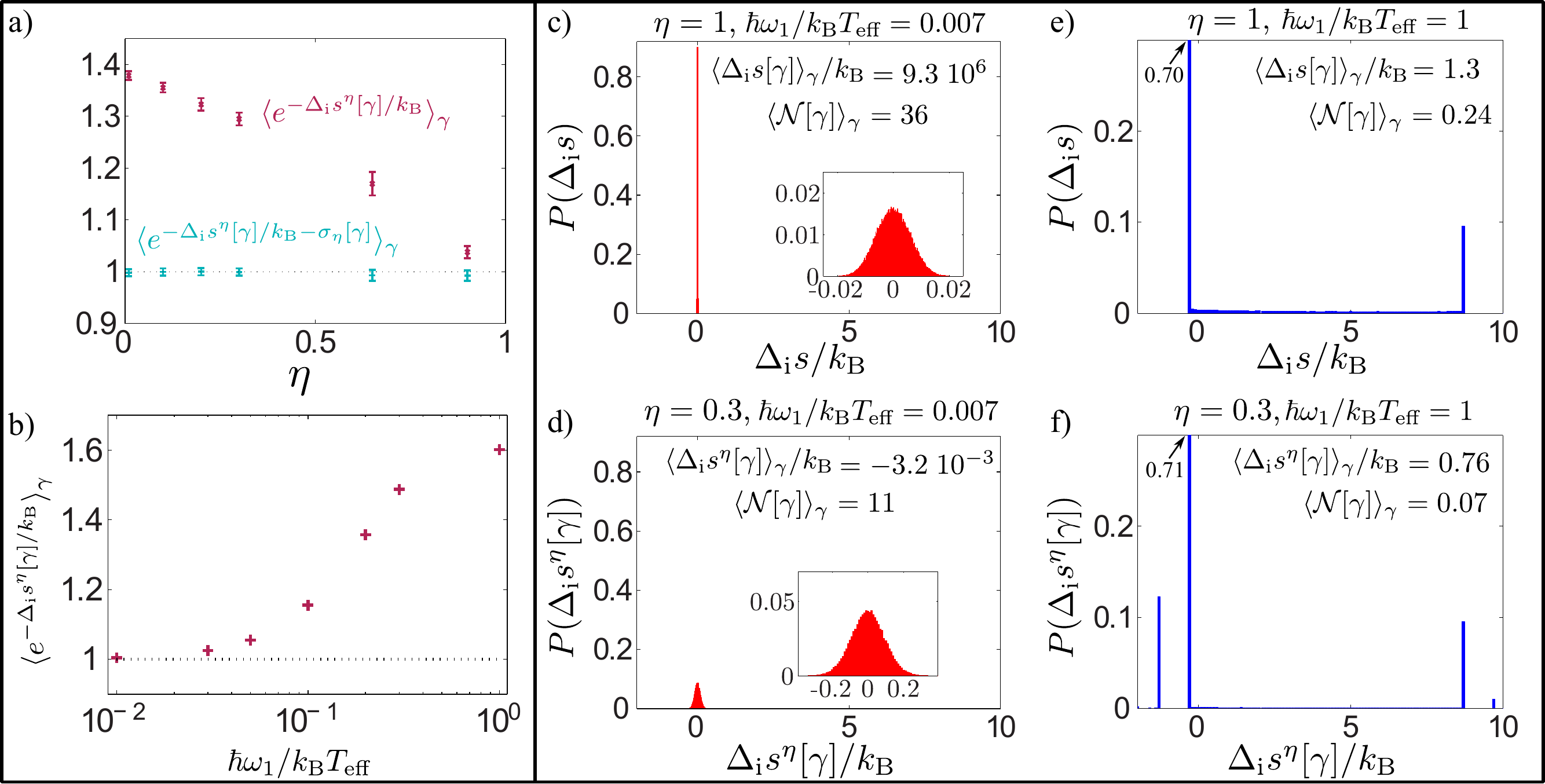}
\end{center}
\caption{Test of Jarzynski equality at finite efficiency, for a drive $H_\text{d}(t) = \hbar \omega_1 \epsilon (t-t_i)/2 \sigma_z$. Purple `x': Measured parameter $\langle e^{-\Delta_\text{i}s^\eta[\gamma]/k_text{B}}\rangle_\gamma$. The magnitude of the deviation (positive for this transformation) increases monotonically when $\eta$ decreases. Blue `+': The plotted parameter included the correction term $\sigma[\gamma]$. We have used $N_\text{traj} = 10^4$ trajectories $\gamma$ and $10^4$ ficitious trajectories $\gamma_F$ for each trajectory. The error bars stand for the standard deviation divided by $\sqrt{N_\text{traj}}$. b) Jarzynski's parameter without correction for $\eta = 0.1$ as a function of the temperature. The error bars are smaller than the marker size. c)-f) Distributions of the entropy production during the transformation for $\eta = 1$ in c),e) and $\eta =  0.3$ in d),f). c) and d) correspond to high temperature ($\beta\omega_1 = 7\; 10^{-3}$) and therefore a large mean number of photons exchanged per trajectory $\moy{{\cal N}[\gamma]}_\gamma = N - \moy{{\cal N}_0[\gamma]}_\gamma$. The inserts show the same distribution with rescaled axes to evidence the similarity of shape. e) and f) correspond to a lower temperature ($\beta\omega_1 = 1$) and a few photon exchanged. {\it Parameters}: For a) $\hbar\omega_1/k_\text{B}T_\text{eff}  = 0.1$, $\gamma t_f = 0.5$, $\epsilon/\gamma = 600$.  For b) and c): $\gamma t_f = 1$, $\epsilon/\gamma = 9$, $N_\text{traj} = 10^5$. \label{fig4}
}
\end{figure*}

\noindent We now extend the definitions of thermodynamic quantities into the finite efficiency regime. For the sake of simplicity, we restrict the study to a driving Hamiltonian of the form $H_\text{d}(t) = (\hbar \omega_1/2) \epsilon (t-t_i) \sigma_z$, such that no quantum heat is exchanged. With the definitions proposed above, the increment of measured classical heat reads:
\bb
\delta Q_\text{cl}^\eta(t_n) =  \left\{\begin{array}{c}
-\hbar\omega_1(t_n),\quad {\cal K}(t_n)=1 \\ 
\hbar\omega_1(t_n),\quad{\cal K}(t_n)=2.
\end{array} \right.
\label{dQR}.
\ee
Despite its apparent similarity with the definition in the perfect efficiency regime, this energy flow does not capture the entire classical heat flow dissipated in the heat bath, as it does not take into account the energy carried by the undetected photons. Therefore, the measured entropy production $\Delta_\text{i}s^\eta[\gamma] = (\Delta U-Q_\text{cl}^\eta[\gamma]-\Delta F)/T_\text{eff}$ does not check Jarzynski equality as evidenced in Fig.\ref{fig4}a) and b). The violation of JE is all the larger as the efficiency $\eta$ is smaller, and JE is recovered in the limit $\eta\to 1$, showing that our definitions for the finite-efficiency case still hold in the perfect efficiency limit. Remarkably, however, it is possible to formulate another equality taking into account the finite efficiency, converging towards the standard JE when $\eta\rightarrow 1$:
\bb
\moy{e^{-\Delta_\text{i}s^\eta[\gamma]/k_\text{B} - \sigma_\eta[\gamma]}}_\gamma = 1. \label{JEeta}
\ee

This equality is reminiscent of JE, but it includes a trajectory-dependent correction term $\sigma_\eta[\gamma]$. We now show that $\sigma_\eta[\gamma]$ can be computed from the measurement record only.\\

The entropy creation along one trajectory $\gamma$  is usually defined as $\Delta_\text{i} s[\gamma] = k_\text{B}\log(P_\text{d}[\gamma]/P_\text{r}[\gamma^\text{r}])$, where $P_\text{d}[\gamma]$ is the probability of a trajectory $\gamma$ in the direct protocol defined by $H_\text{d}(t)$:
\bb
P_\text{d}[\gamma] = p_\text{i} \text{Tr}\left\{\ket{f}\bra{f}\left(\overrightarrow{\prod_{n=1}^N} {\cal E}_{{\cal K}_\gamma(t_n)}^{(\eta)}\right)\big[\ket{i}\bra{i}\big]\right\},\label{Pdeta}
\ee
\noindent We have introduced $p_\text{i} = Z^{-1}_\text{i}e^{-U_\text{i}/k_\text{B}T_\text{eff}}$ the probability of drawing the initial pure state $\ket{i}\bra{i}$ of internal energy $U_\text{i}$ in the initial thermal distribution. The arrow indicates the order in which the sequence of super-operators is applied.  $P_\text{r}[\gamma^\text{r}]$ is the probability of the time-reversed trajectory $\gamma^\text{r}$ corresponding to $\gamma$. It is generated by the time-reversed drive $H_\text{d}(t_\text{f}-t)$ and the reversed sequence of super-operators ${\cal E}^{(\eta),r}_{{\cal K}_\gamma(t_n)}$ \cite{Manzano15,Crooks08,Elouard16}, such that:

\bb
P_\text{r}[\gamma^\text{r}] = p_\text{f}  \text{Tr}\left\{\ket{i}\bra{i}\left(\overleftarrow{\prod_{n=1}^N} {\cal E}_{{\cal K}_\gamma(t_n)}^{(\eta),\text{r}}\right)\big[\ket{f}\bra{f}\big]\right\},\label{Preta}
\ee
\noindent where $p_\text{f} = Z^{-1}_\text{f}e^{-U_\text{f}/k_\text{B}T_\text{eff}}$ the probability of drawing the final pure state $\ket{f}\bra{f}$ of internal energy $U_\text{f}$ in the final thermal distribution.

When $\eta=1$, the time-reversed operators at time $t_n$ reduce to
\bb 
{\cal E}^{(1),\text{r}}_1[\rho_\gamma](t_n) &=& e^{-\hbar \omega_1(t_n)/k_\text{B}T_\text{eff}}M_1^\dagger \rho_\gamma M_1 \label{Er1}\\
{\cal E}^{(1),\text{r}}_2[\rho_\gamma] (t_n)&=& e^{\hbar \omega_1(t_n)/k_\text{B}T_\text{eff}}M_2^\dagger \rho_\gamma M_2 \label{Er2}\\
{\cal E}^{(1),\text{r}}_0[\rho_\gamma](t_n) &=& M_0^\dagger \rho_\gamma M_0.\label{Er0}
\ee

\noindent By inserting these expressions in Eq.\eqref{Preta}, we express the ratio $P_\text{r}[\gamma^\text{r}]/P_\text{d}[\gamma] = e^{-(\Delta U[\gamma] - \Delta F)/k_\text{B}T_\text{eff} }e^{Q_\text{cl}[\gamma]/k_\text{B}T_\text{eff} }$, which is JE. \\

When $\eta < 1$, the ratio $P_\text{r}[\gamma^\text{r}]/P_\text{d}[\gamma]$ is modified. It is useful to decompose ${\cal E}^{(\eta)}_0$ into three super-operators conserving purity: 
\bb
{\cal E}^{(\eta)}_0 = {\cal E}^{(\eta)}_{00}+{\cal E}^{(\eta)}_{01}+{\cal E}^{(\eta)}_{02},\label{decomposition}
\ee 
with 
\bb
{\cal E}^{(\eta)}_{01}[\rho_\gamma] &=& (1-\eta){\cal E}^{(1)}_1[\rho_\gamma], \nonumber\\
{\cal E}^{(\eta)}_{02}[\rho_\gamma] &=& (1-\eta){\cal E}^{(1)}_2[\rho_\gamma], \nonumber\\
{\cal E}^{(\eta)}_{00}[\rho_\gamma] &=& {\cal E}^{(1)}_0[\rho_\gamma].\label{E0i}
\ee

\noindent From this decomposition, we generate a set ${\cal F}[\gamma]$ of fictitious trajectories $\gamma_F$ of pure states, compatible with $\gamma$: a trajectory $\gamma_F \in  {\cal F}[\gamma]$ is built by applying a sequence of super-operators $\{{\cal E}_{{\cal K}_{\gamma_F}(t_n)}\}_n$, with ${\cal K}_{\gamma_F}(t_n) = {\cal K}_{\gamma}(t_n)$ if ${\cal K}_{\gamma}(t_n)$ belongs to $\{1,2\}$ (i.e. when a photon has been detected), and ${\cal K}_{\gamma_F}(t_n) \in \{00,01,02\}$ when ${\cal K}_{\gamma}(t_n)$ is zero (no photon detected). ${\cal F}[\gamma]$ thus contains $3^{{\cal N}_0[\gamma]}$ fictitious trajectories, where ${\cal N}_0[\gamma]$ is the number of time-steps during which no photon is detected. 

The perfect-efficiency fictitious trajectories are interesting because they can be time-reversed using the rules Eq.\eqref{Er1}-\eqref{Er0}, while their probability distribution checks $P_\text{d}[\gamma] = \sum_{\gamma_F\in \mathcal{F}[\gamma] } P_\text{d}[\gamma_F]$. Eventually, the following equality is derived:
\bb
\sum_{\gamma} P_\text{r}[\gamma^r] &=& 1\nonumber\\
&=&  \sum_{\gamma} \sum_{{\cal F [\gamma]}}  P_\text{r}[\gamma_F^r]\nonumber\\
&=& \sum_{\gamma} \sum_{{\cal F [\gamma]}} P_\text{d}[\gamma_F] e^{-(\Delta U[\gamma_F]-\Delta F - Q_\text{cl}[\gamma_F])/k_\text{B}T_\text{eff}}\nonumber\\
&=& \sum_{\gamma} P_\text{d}[\gamma] e^{-(\Delta U[\gamma_R]-\Delta F - Q_\text{cl}^{\eta}[\gamma])/k_\text{B}T_\text{eff}}\nonumber\\
&&\quad\times\sum_{{\cal F [\gamma]}} P_\text{d}[\gamma_F\vert \gamma] e^{( Q_\text{cl}^{\eta}[\gamma]- Q_\text{cl}[\gamma_F])/k_\text{B}T_\text{eff}}.\nonumber\\\label{Pdeta2}
\ee
\noindent We have introduced the conditional probability $P_\text{d}[\gamma_F\vert \gamma] = P_\text{d}[\gamma_F]/P_\text{d}[\gamma]$ of the fictitious trajectory $\gamma_F\in {\cal F}[\gamma]$, given the detected trajectory $\gamma$. In order to find Eq.\eqref{JEeta}, we now define: 

\bb
\sigma_\eta[\gamma] = -\log \sum_{{\cal F}[\gamma]} P(\gamma_F\vert\gamma) e^{(Q_\text{cl}[\gamma_F]-Q_\text{cl}^\eta[\gamma])/k_\text{B}T_\text{eff}}.
\ee

\noindent $\sigma_\eta[\gamma]$ can be numerically computed, based solely on the measurement outcomes. The computation involves the simulation of a sample of the trajectories $\gamma_F \in {\cal F}[\gamma]$ for each trajectory $\gamma$. The corrected expression $\langle e^{-\Delta_\text{i}s[\gamma]/k_\text{B} -\sigma_\eta[\gamma]} \rangle_\gamma$ is plotted in Fig.\ref{fig4}a): JE is verified, showing that the experimental demonstration of a FT in a realistic setup is within reach.\\

Our results explicit a new quantum-classical border in the thermodynamic framework, which is illustrated in Fig.\ref{fig4}(b to f) where the deviation from JE is studied as a function of the reservoir temperature. In the quasi-static limit characterized by $\epsilon \ll \Gamma_- $ (high temperature case), many photons are exchanged and the system is allowed to thermalize before any significant work is done: the transformation is reversible. We see from Fig.\ref{fig4}c) and d) that in this limit, the effect of detection inefficiency in the entropy production is merely to widen the spread of the distribution but conserving its Gaussian shape around zero.  The statistical properties of the distribution are not affected, even if many photons are missed, and JE is verified (Fig.\ref{fig4}b). This corresponds to a semi-classical situation where, even though the exchange of heat is quantized, the overall thermodynamic behavior of the system is equivalent to a classical ensemble. In this limit, there is no significant information loss due to the undetected photons.

On the other hand, as soon as work and heat are exchanged at similar rates $\epsilon \sim \Gamma_-$, the system is brought out of equilibrium and the transformation is irreversible. In this limit, few photons are exchanged before a non negligible amount of work is done on the system. Each missed photon represents, then, a significant information loss,  such that the measured distribution of entropy production is severely affected (see Fig.\ref{fig4}e) and f)). Such information loss is quantifiable by the violation of the non-corrected JE, and is a direct consequence of the quantization of heat exchanges.\\

\section{Conclusion}
\label{Conclusion}
We presented a realistic setup based on a driven three-level atom allowing to simulate a qubit in equilibrium with an engineered reservoir, giving full access to the distribution of the heat dissipated by the qubit. We exploited this scheme to characterize irreversibility in the case of a coherently driven qubit, coupled to a thermal bath: we evidenced genuinely quantum contributions to the entropy production, and showed these contributions do have some energetic counterpart. In a second part, we took into account finite detection efficiency and evidenced deviations from Jarzynski equality that can be tested and corrected. We derived a modified equality, involving only quantities computed from the finite-efficiency measurement record. 
This work opens avenues for the experimental verification of fluctuation theorems in quantum open systems. Owing to the versatility of the scheme, various reservoirs could be simulated, including non-thermal ones. Furthermore, quantitative relations between entropy production and energy fluctuations are the basis for energetic bounds for classical computation \cite{Landauer61,Bennett82,Faist15}. In this work we study situations where equivalent relations can be derived in the quantum regime, providing new tools to explore energetic bounds for quantum information processing.\\

\section*{Acknowledgments}
MFS acknowledges CNPq (Project 305384/2015-5). NKB thanks the support from the Brazilian agency CAPES. This work was supported by the Fondation Nanosciences of Grenoble under the Chair of Excellence "EPOCA", by the ANR under the grant 13-JCJC-INCAL and by the COST network MP1209 "Thermodynamics in the quantum regime".

\section*{Appendix}

\subsection{Simulating a heat bath with an engineered environment}

The rates $\Gamma_\pm$ induce the same dynamics as a heat bath, if and only if
$$\dot P_e = -\Gamma_0(2\bar n[\omega_1(t)]+1)\left(P_e(t)-\dfrac{\bar n[\omega_1(t)]}{2\bar n[\omega_1(t)] +1}\right).$$ 
\noindent We have introduced the population of the excited qubit's state $P_e = (\text{Tr}[\rho(t) \ket{e}\bra{e}]$, the mean number of thermal photons in the effective bath $\bar n [\omega] = (e^{\hbar \omega/k_\text{B} T_\text{eff}}-1)^{-1}$ and $\Gamma_0$ the spontaneous emission rate of the qubit. The dynamics induced by the engineered environment is:
$$
\dot P_e = -(\Gamma_++\Gamma_-)\left(P_e(t)-\dfrac{\Gamma_+}{\Gamma_++\Gamma_-}\right).
$$ 
By identification, we find the conditions:

\bb
\Omega(t) &=& \Gamma_0 \sqrt{(\bar n[\omega_1(t)]+1)\bar n[\omega_1(t)]}\label{PrOmega}\\
\Gamma_-(t) &=& \Gamma_0 (\bar n[\omega_1(t)]+1) \label{PrGamma},
\ee

\noindent Eq.\eqref{PrOmega} can be fulfilled by tuning accordingly the intensity of the laser drive. Eq.\eqref{PrGamma} requires to tune the incoherent rate of transition from $\ket{e}$ to $\ket{g}$. This task can be performed by embedding the three level atom in a quasi-resonant optical cavity and tuning the frequency between state $\ket{e}$ and $\ket{g}$ e.g. with a non-resonant laser (Stark effect) according to a protocol $\omega_1'(t)$ designed to fulfill Eq.\eqref{PrGamma}. Note that the implemented protocol $\omega_1'(t)$ is in general different from the target protocol $\omega_1(t)$ which defines the constraints Eq.\eqref{PrOmega}-\eqref{PrGamma} on the qubit dynamics. \\

\subsection{Expression of the quantum heat}

For a general driving Hamiltonian $H_\text{d}(t) = \delta(t)\sigma_z + \mu(t)\sigma + \mu^*(t)\sigma^\dagger$, the quantum heat $\delta Q_\text{q}(t_n)$ exchanged when a jump takes place reads (for $\eta = 1$):

\bb
\delta Q_\text{q} &=& \left\{\begin{array}{c}  \hbar\omega_1(t)P_g(t) - 2\text{Re}\big(\mu(t)\moy{\sigma(t)},\quad{\cal K}_\gamma(t_n) = 1\nonumber\\
-\hbar\omega_1(t)P_e(t)- 2\text{Re}\big(\mu(t)\moy{\sigma(t)},\quad{\cal K}_\gamma(t_n) = 2  \end{array} \right.
\ee

\noindent Between two photon detections, the quantum heat increment reads:

\bb
\delta Q_\text{q}(t_n) &=& - (\Gamma_--\Gamma_+)dt \text{Re}(\moy{\sigma})(\hbar\mu(t_n)P_g(t_n)-\hbar\mu^*(t_n)P_e(t_n)) \nonumber\\
&&- \hbar\omega_1(t_n)(\Gamma_--\Gamma_+)dt P_e(t_n)P_g(t_n) ,\;\;{\cal K}_\gamma(t_n) = 0\;\; \label{C:dQq}
\ee

Note that $\delta Q_\text{q}(t_n)$ is zero when the qubit state has no coherence in $\ket{e}-\ket{g}$ basis, such that $\moy{\sigma}=\moy{\sigma^\dagger} =0$, $(P_e(t_n),P_g(t_n)) = (1,0)$ (resp. $(0,1)$) when ${\cal K}_\gamma(t_n) = 1$ (resp. ${\cal K}_\gamma(t_n) = 2$): the qubit has zero probability to absorb a photon at time $t$ when $P_e(t_n) = 1$, and zero probability to emit a photon when $P_g(t_n)= 1$.

\bibliography{FT}{}

\end{document}